\documentclass[journal=jacsat,manuscript=communication,layout=twocolumn]{achemso}

\usepackage{chemformula} 
\usepackage[T1]{fontenc} 

\usepackage{graphicx} 
\graphicspath{ {./images/} }

\usepackage[rightcaption]{sidecap}

\usepackage{wrapfig}

\SectionsOn

\author{Tim R. Eichhorn}
\altaffiliation{Contributed equally to this work}
\author{Anna J. Parker}
\altaffiliation{Contributed equally to this work}
\author{Felix Josten}
\author{Christoph M\"{u}ller}
\author{Jochen Scheuer}

\author{Jakob M. Steiner}
\affiliation[NVision GmbH]
{NVision Imaging Technologies GmbH, 89081 Ulm, Germany}
\alsoaffiliation[PSI]{Paul Scherrer Institute, 5232 Villigen PSI, Switzerland}

\author{Martin Gierse}
\affiliation[NVision GmbH]
{NVision Imaging Technologies GmbH, 89081 Ulm, Germany}
\alsoaffiliation[UniUlm]
{Institute for Quantum Optics, Ulm University, 89081 Ulm, Germany}

\author{Jonas Handwerker}
\author{Michael Keim}
\author{Sebastian Lucas}

\author{Mohammad Usman Qureshi}
\affiliation[NVision GmbH]
{NVision Imaging Technologies GmbH, 89081 Ulm, Germany}

\author{Alastair Marshall}
\affiliation[NVision GmbH]
{NVision Imaging Technologies GmbH, 89081 Ulm, Germany}
\alsoaffiliation[UniUlm]
{Institute for Quantum Optics, Ulm University, 89081 Ulm, Germany}

\author{Alon Salhov}
\affiliation[NVision GmbH]
{NVision Imaging Technologies GmbH, 89081 Ulm, Germany}

\author{Yifan Quan}
\affiliation[PSI]{Paul Scherrer Institute, 5232 Villigen PSI, Switzerland}

\author{Jan Binder}
\author{Kay Jahnke}
\affiliation[NVision GmbH]
{NVision Imaging Technologies GmbH, 89081 Ulm, Germany}

\author{Philipp Neumann}
\author{Stephan Knecht}
\author{John W. Blanchard}
\affiliation[NVision GmbH]
{NVision Imaging Technologies GmbH, 89081 Ulm, Germany}

\author{Martin B. Plenio}
\affiliation[UniUlm]
{Institute for Theoretical Physics, Ulm University, 89081 Ulm, Germany}
\alsoaffiliation{Center for Integrated Quantum Science and Technology, Ulm University, 89081 Ulm, Germany}

\author{Fedor Jelezko}
\affiliation[UniUlm]
{Institute for Quantum Optics, Ulm University, 89081 Ulm, Germany}
\alsoaffiliation{Center for Integrated Quantum Science and Technology, Ulm University, 89081 Ulm, Germany}

\author{Lyndon Emsley}
\affiliation[EPFL]
{Institut des Sciences et Ing\'{e}nierie Chimiques, \'{E}cole Polytechnique F\'{e}d\'{e}rale de Lausanne, CH-1015 Lausanne, Switzerland}

\author{Christophoros C. Vassiliou}
\affiliation[NVision GmbH]
{NVision Imaging Technologies GmbH, 89081 Ulm, Germany}

\author{Patrick Hautle}
\affiliation[PSI]{Paul Scherrer Institute, 5232 Villigen PSI, Switzerland}

\author{Ilai Schwartz}
\affiliation[NVision GmbH]
{NVision Imaging Technologies GmbH, 89081 Ulm, Germany}

\email{ilai@nvision-imaging.com}

\title[Optically polarized crystals for NMR]
  {Hyperpolarized solution-state NMR spectroscopy with optically polarized crystals}

\begin{document}

\begin{abstract}

Nuclear spin hyperpolarization provides a promising route to overcome the challenges imposed by the limited sensitivity of nuclear magnetic resonance.
Here we demonstrate that dissolution of spin-polarized pentacene-doped naphthalene crystals enables transfer of polarization to target molecules via intermolecular cross relaxation at room temperature and moderate magnetic fields (1.45$\,$T).
This makes it possible to exploit the high spin polarization of optically polarized crystals while 
mitigating the challenges of its transfer to external nuclei, particularly of the large distances and prohibitively weak coupling between source and target nuclei across solid-solid or solid-liquid interfaces. 
With this method, here we inject the highly polarized mixture into a benchtop NMR spectrometer and observe the polarization dynamics for target $^1$H nuclei. 
Although the spectra are radiation damped due to the high naphthalene magnetization, we describe a procedure to process the data in order to obtain more conventional NMR spectra, and extract the target nuclei polarization.
With the entire process occurring on a timescale of one minute, we observe NMR signals enhanced by factors between -200 and -1730 at 1.45$\,$T for a range of small molecules.

\end{abstract}

\section{Introduction}
High-field NMR spectroscopy is the cornerstone analytical technique in the chemical sciences today, with applications ranging from the determination of chemical structures in synthetic intermediates to the determination of the atomic-level structure and dynamics in proteins and nucleic acids. However, the Achilles' heel of NMR is its limited sensitivity 
(due to a combination of the minute size of nuclear magnetic moments and the correspondingly small polarization at thermal equilibrium),
which means that many experiments are time or material consuming and various potential applications are inaccessible. 
As a result, substantial effort has been put into increasing NMR sensitivity, notably through the use of higher magnetic fields and optimized detection systems \cite{Styles1984,Luchinat2021}. 
An alternative approach to increase NMR sensitivity is through hyperpolarization \cite{Overhauser:1953, Carver:1956,Ardenkjaer:2003}, where the nuclear spins are polarized to a level significantly greater than thermal equilibrium.
Hyperpolarization techniques offer promising prospective solutions for observing low-gamma nuclei or low-concentration analytes, for example.

Cryogenic dynamic nuclear polarization (DNP) \cite{Abragam1978,wenckebach2016essentials} is arguably the most general hyperpolarization method, and is used to great success in materials science \cite{michaelis2019handbook,perras2007growing,Rossini2013}, nuclear physics \cite{niinikoski2020}, and medical imaging applications \cite{Ardenkjaer:2003, Nelson:2013, Wang:2019}. 

The instrumentation requirements for DNP are significant, however, because generating substantial equilibrium electron spin polarization simultaneously requires high magnetic fields (typically several T) and cryogenic temperatures \cite{Rosay2010}.
Alternatively, it is possible to perform DNP in the solution state (so-called Overhauser DNP) near to room temperature, but signal enhancements are typically limited to a factor of 5-20\cite{Ebert2012,Griesinger2012,Denysenkov2017}.
Significant progress has been made with other methods, such as those based on parahydrogen-induced polarization (PHIP) \cite{BowersWeitekamp,Green2012PHIP,Knecht2021} or spin-exchange optical pumping (SEOP) of noble gases \cite{WalkerHapper,Nikolaou2013}, but they are currently only applicable to chemically specific systems. 

As a result, a broadly accessible, general approach for signal enhancement in solution-state NMR remains elusive.

\begin{figure*}
  \centering 
  \includegraphics[width=0.75\textwidth]{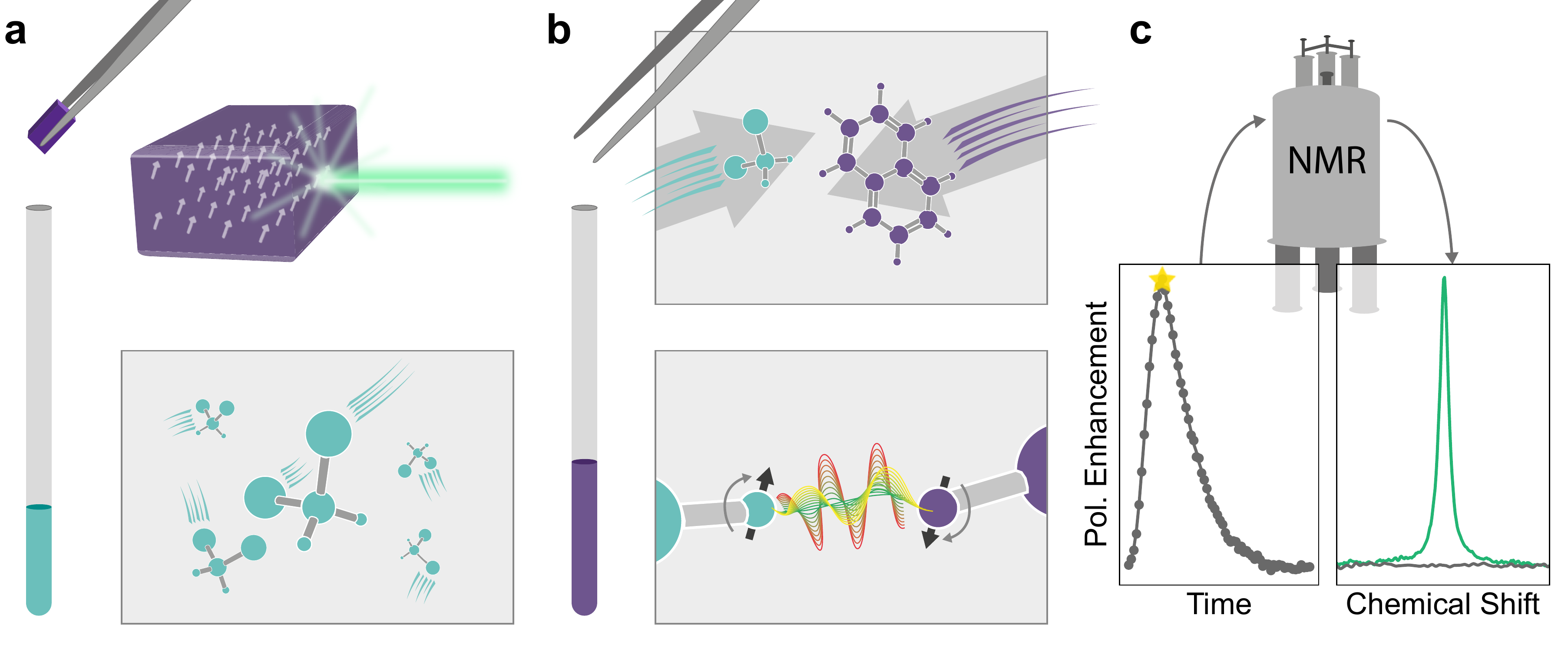}
  \caption{Illustration of the workflow using optically polarized organic crystals for signal amplification in solution state NMR spectroscopy: 
  (\textbf{a}) Triplet-DNP of pentacene-doped naphthalene single crystals with laser excitation yields high-density polarization sources with $20-25\%$ $^1$H spin polarization. 
  (\textbf{b}) Dissolution of the naphthalene crystal with target analyte at room temperature initiates the intermolecular nuclear Overhauser effect (NOE) 
  which leads to transfer of polarization that (\textbf{c}) can be utilized in a high-field NMR spectrometer.
  }
  \label{fgr:fig1}
\end{figure*}

Polarization via optically initialized electron triplet states, e.g., in organic molecular crystals\cite{Deimling:1980,Henstra:1990,Iinuma:2000} or in NV$^-$ defect centers in diamond\cite{london2013detecting}, has several attractive features. 
The electron polarization does not depend on the thermodynamic equilibrium, thereby enabling up to tens of percent nuclear-spin polarization at room temperature and low magnetic fields (< 1 T)\cite{Tateishi:2014,King:2015} .
The polarization rate can be controlled by the optical pumping rate, 
and in the case of photoexcited triplet-state electrons, the relaxation time of the nuclear spins following polarization can be exceedingly long, as the paramagnetic electrons are only present under laser irradiation. For example, $^1$H nuclei in pentacene-doped naphthalene crystals have been polarized to a record 80\% , with a lifetime of approx. 50 hours at liquid nitrogen temperatures\cite{Quan:2019}, long enough to be transported to remote facilities. 

The outstanding challenge for NMR spectroscopy is the transfer of this polarization from the crystal to molecules of interest. Various strategies of applying triplet-DNP to hyperpolarization in solution-state NMR have been studied, including metal-organic frameworks\cite{Fujiwara:2018}, porphyrins\cite{Hamachi:2021} and water-soluble polarizating agents\cite{Kouno:2020}. 
Furthermore, intensive research efforts have focused on the properties of the triplet-DNP source material, uncovering a range of interrelated problems, including surface material quality \cite{Chu:2014}, 
properties of the optically polarized defects (especially when increasing their density) \cite{eichhorn2019optimizing}, and producing crystals with sufficient surface area to polarize measurable volumes of material \cite{Waddington2017, Broadway2018}. 

Here we show that dissolving hyperpolarized naphthalene crystals in solutions of target analytes leads to polarization transfer via the intermolecular nuclear Overhauser effect (NOE), providing NMR signal enhancements across a range of substrates.
The overall process is illustrated in Fig.\,\ref{fgr:fig1}. 

The polarization transfer occurs on the timescale of less than one minute, 
does not require cryogenics, 
and results in high-resolution NMR spectra due to the absence of paramagnetic contaminants. 
Compared to other hyperpolarization technologies, the polarization-transfer system is fully automated, operates at room temperature, and is mounted on an existing NMR spectrometer, requiring no additional lab/bench space. 

We demonstrate the method for a number of small molecules and molecular mixtures, regularly achieving signal enhancements greater than a factor of -200 and up to a factor of -1730 (corresponding to 0.86\% polarization) for the benchmark molecule propargyl acetate at 1.45 T .

\section{Results}

\begin{figure*}
  \centering 
  \includegraphics[width=0.8\textwidth]{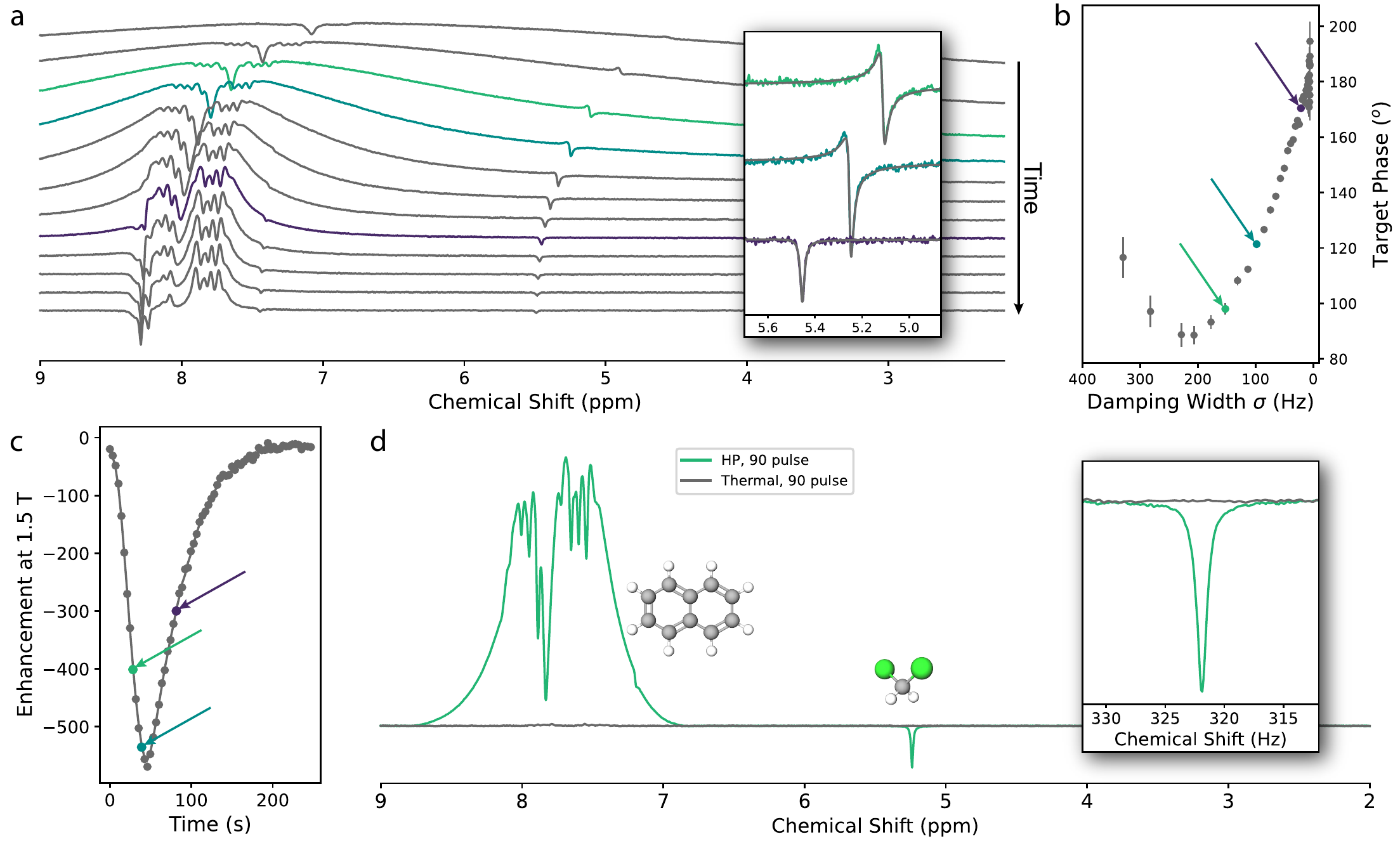}
  \caption{NMR phenomena in solutions containing hyperpolarized naphthalene and dichloromethane. 
  \textbf{a}) Time series of $^1$H NMR spectra following injection of the hyperpolarized solution into the magnet.
  Spectra shown are separated by 10\,s.
  The resonances shift over time, with the total shift decaying to zero exponentially over time. The inset shows the dichloromethane signal at 20, 30, and 70\,s to demonstrate the effect of the radiation-damping field on the phase of the target $^1$H resonance.
  \textbf{b}) The relative phase of the target peak correlates with the width of the damped naphthalene spectrum $\sigma_{rd}$, starting at around 90$^\circ$ and evolving to 180$^\circ$, as expected for signal enhancement by intermolecular NOE.
  \textbf{c}, Enhancement of the target polarization over time after injection into the spectrometer.
  Based on the polarization build-up curve, the optimum polarization enhancement for DCM is achieved approximately 35\,s after injection.
  \textbf{d} Hyperpolarized $^1$H NMR spectrum resulting from a subsequent experiment featuring a 90$^\circ$ pulse applied 35\,s after sample injection.
  The inset shows the dichloromethane signal to show the amplification relative to the thermal signal as well as to emphasize that the resonance width is still roughly 1\,Hz,
a feature not always possible with hyperpolarization techniques.}
  \label{fgr:dataanalysis}
\end{figure*}

As depicted in Fig.~\ref{fgr:fig1}, the first step of the hyperpolarization procedure is to optically polarize a single crystal of pentacene-doped naphthalene to 20-25\% $^1$H polarization (see Methods).
The polarized crystal is then transported at room temperature in a handheld permanent magnet assembly to the polarization-transfer device, where it is inserted into a dissolution vessel under inert atmosphere, crushed, and dissolved into a solution of interest.
Finally, the sample is injected into a 1.45\,T benchtop spectrometer for measurement.
For time-dependent studies, small-angle ( or 2.0$^\circ$) rf pulses are applied every 3.3\,s in order to monitor source and target signals.
Alternatively, in order to acquire the maximum signal intensity, a single $90^\circ$ pulse can be applied after waiting for polarization transfer.

\subsubsection{Analysis and interpretation of hyperpolarized spectra}
Figure~\ref{fgr:dataanalysis}a shows a series of NMR spectra at 10\,s intervals (spectra were acquired with 2.0$^\circ$ rf pulses every 3.3\,s -- for clarity, only every three spectra are shown) following injection of the hyperpolarized solution into the spectrometer. The sample considered in this example contains approximately 1.8\,M naphthalene (20\%v/v) and\,100 mM dichloromethane in CDCl$_3$.
The spectra show a number of striking features related to the large nuclear magnetization:

First, there is a significant drift of the resonances due to the magnetic field produced by the sample, shown clearly by the change in position of the dichloromethane and broadened naphthalene resonances over time.
For a long cylinder magnetized transverse to its symmetry axis, the magnetic field within a small spherical cavity inside the cylinder is $\mathbf{B_M}=\mu_0 \mathbf{M}/2-2\mu_0\mathbf{M}/3=-\mu_0\mathbf{M}/6$,
with nuclear magnetization $\mathbf{M}=N P \hbar \gamma_I \mathbf{I}$, where $N$ is the number density of nuclear spins, $P$ is the spin polarization, $\hbar$ is the reduced Planck constant, and $\gamma_I$ is the nuclear gyromagnetic moment of spins $\mathbf{I}$.
The nuclear spin polarization can therefore be determined based on the frequency shift, $\delta$ (expressed in ppm):
\begin{equation}
    P = -\frac{12 B_0}{\mu_0 \hbar \gamma_I N}\delta\times 10^{-6} .
    \label{eqn:delta_shift}
\end{equation}
The magnetic field of the benchtop NMR spectrometer is 1.45\,T, so a shift of -1.00\,ppm corresponds to 5.73\% $^1$H polarization.

Second, the spectra are dominated by a large broad peak centered at the naphthalene resonance frequency.
The broadening is due to back action of the detection circuit on the spins, which produces negative feedback, referred to in NMR as ``radiation damping''\cite{Bloembergen1954,Augustine2002}.
Figure\,\ref{fgr:dataanalysis}a shows the spectrum acquired 35\,s after injection into the spectrometer.
Because the radiation damping of the naphthalene signal is so dramatic, it is easily distinguished from other resonances, and can thus be subtracted (see below).
It is also worth noting that \emph{positive} feedback is possible if the source is oppositely polarized, as in the so-called ``RASER'' (radio amplification by stimulated emission of radiation) effect \cite{Suefke2017,Appelt2021} -- while this involves some interesting spin physics, it is generally detrimental for practical applications, therefore we ensure that the source polarization is oriented along the field.

The large naphthalene signals are also strong enough to affect other spin species.
The relatively short-lived magnetic fields induced by the precessing spins and detector back action couple to other spins and can be thought of as a weak rf pulse, inducing a frequency-dependent phase shift on other resonances.
To zeroth order, these fields are directly related to the radiation-damped signals measured by the spectrometer.
As such, the phase of target resonances are approximately linearly dependent on the magnitude of the naphthalene signal at the target resonance frequency. 
The fit of the broadened naphthalene signal can therefore be multiplied by a constant to define a frequency-dependent phase correction.

The width of the radiation-damped signal is related to the magnitude of the sample magnetization and the intensity of the radiation damping itself, so there is also a relationship between the phase of the target resonance and the width of the radiation-damped signal, as shown in Fig.\,\ref{fgr:dataanalysis}b.

By understanding these somewhat exotic effects, it becomes possible to separate them from the rest of the spectrum in order to recover a more conventional 1D NMR spectrum for the hyperpolarized analyte(s) via the following procedure:\\[6pt]
\noindent \hspace*{\parindent}1. The damped naphthalene signal is fit to a complex Lorentzian lineshape,\footnote{Ideally, a sum of Lorentzians, in order to account for higher-order effects of radiation damping.} which is treated as background and subtracted. 
Any residual distortion of the baseline can be corrected using a moving average or other standard techniques.\\
\hspace*{\parindent}2. The magnetic field drift is then estimated by fitting one or more target signals to complex Lorentzian lineshapes. 
The extracted center frequencies are then fit to a decaying exponential, and this function defines the shifts necessary to bring the spectra into alignment. 
In some exceptional cases where the fitting procedure was more difficult, the naphthalene center frequency was used for an approximate correction, followed by manual adjustments to align the spectra. \\
\hspace*{\parindent}3. A nonlinear phase correction is applied based on the fit of the damped naphthalene signal, followed by standard zero- and first-order phase adjustment.\\[6pt]
\hspace*{\parindent}Signal enhancement is defined as the quotient of the integral of the hyperpolarized resonance divided by the integral of the thermally polarized resonance.
These integrals and uncertainties are obtained by fitting the signals to a complex Lorentzian lineshape. 
Because the hyperpolarized spectra--acquired with small-angle pulses--are compared to thermally polarized spectra acquired with a 90$^\circ$ pulse, the signal enhancements are converted to polarization enhancements by multiplying by a calibration factor (40.8 for 1.4$^\circ$, 29.2 for 2.0$^\circ$ excitations), which was measured on the spectrometer.  

\begin{figure*}
  \centering 
 \includegraphics[width=0.8\textwidth]{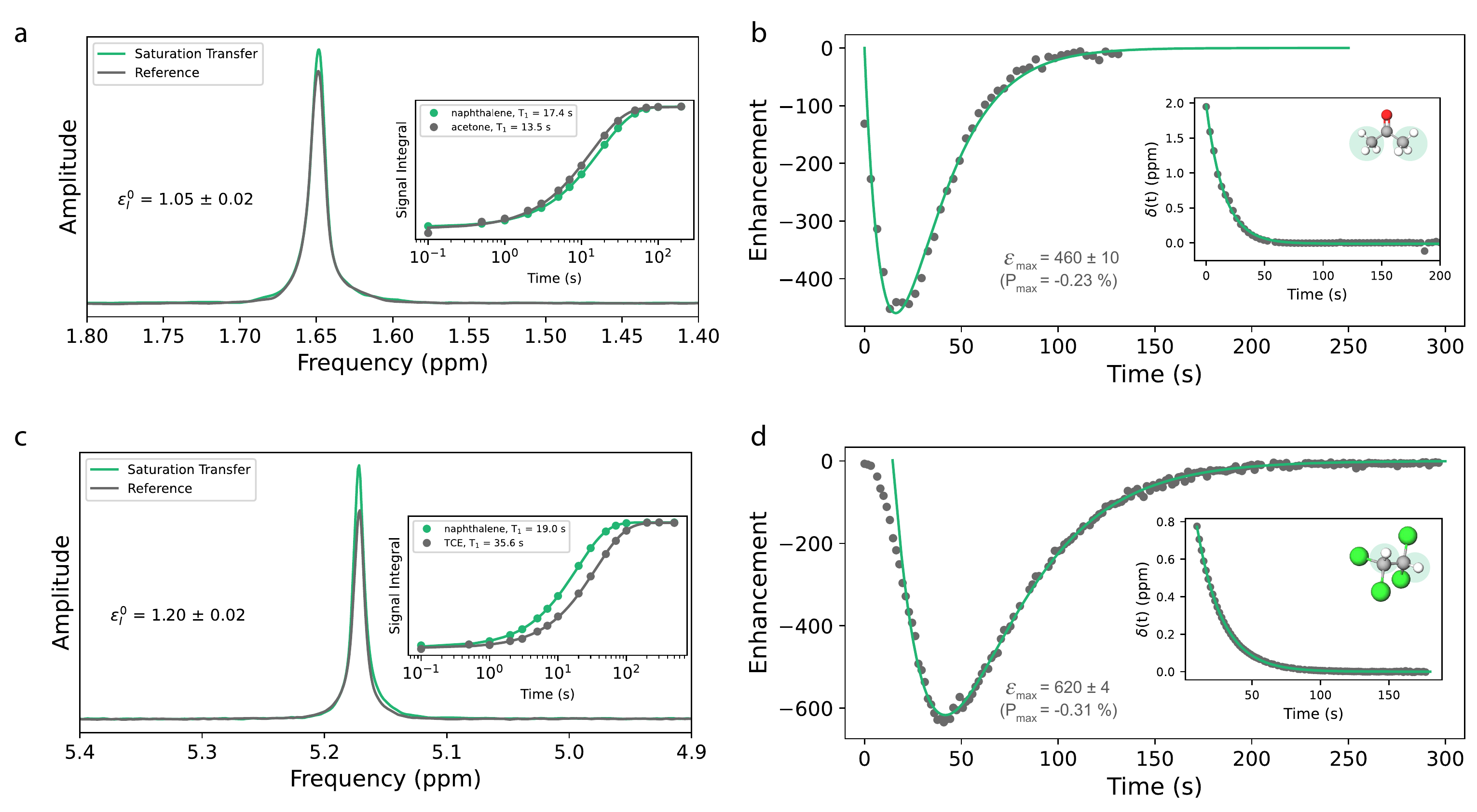}
 \caption{
 Observation of intermolecular NOE in solutions containing 20 \%(v/v) naphthalene-h$_8$ and low concentrations of either acetone (\textbf{a,b}) or TCE in CDCl$_3$ (\textbf{c,d}).
 (\textbf{a,c}) Steady-state saturation transfer with thermally polarized naphthalene gives intermolecular NOE enhancements of $\epsilon_{\rm{I}}^0 (\text{acetone}) = (1.05 \pm 0.02)$ (\textbf{a}) and $\epsilon_{\rm{I}}^0 (\rm{TCE}) = (1.20 \pm 0.02)$ (\textbf{c}). 
 Using the corresponding T$_1$ relaxation times of the target spins (insets of \textbf{a,c} with monoexponential fits), the cross-relaxation rates can be calculated as $\sigma(\rm{acetone}) = (3.7 \pm 1.5) \times 10^{-3} \, \text{s}^{-1}$ and $\sigma(\text{TCE}) = (5.6 \pm 0.6) \times 10^{-3} \, \text{s}^{-1}$. 
  (\textbf{b,d}) Transient polarization transfer with hyperpolarized naphthalene to TCE (\textbf{b}) and acetone (\textbf{d}).
The source polarization at the beginning of the NOE buildup is calculated from the observed resonance frequency shifts (insets \textbf{b,d}) and eq.\,\ref{eqn:delta_shift}, yielding $(11.18 \pm 0.06)\,\%$ for the acetone experiment and $(4.45 \pm 0.01)\,\%$ for the TCE experiment.
Fitting to eq.\,\ref{eqn:NOE_transient_modified} yields cross-relaxation rates $\sigma=(3.62 \pm 0.04)\times 10^{-3}\,\rm{s}^{-1}$ for acetone (\textbf{b}) and 
$\sigma=(6.24 \pm 0.04)\times 10^{-3}\,\rm{s}^{-1}$ for TCE (\textbf{d}), consistent with the thermal measurements.} 
  \label{fgr:fig2}
\end{figure*}

The results of the procedure are shown in Figure \ref{fgr:dataanalysis}c,d, where c shows the enhancement curves estimated for each $^1$H site and d compares a 90$^\circ$ excitation of the final hyperpolarized dichloromethane signal (measured after ca. 35 s) to a single-shot thermal spectrum.

\subsubsection{NOE mechanism of intermolecular polarization transfer in liquids}

We identify the polarization transfer mechanism as resulting from an intermolecular nuclear Overhauser effect (NOE) in liquids, which can be understood as a transfer of magnetization between nuclear spins on molecules in close proximity, induced by the stochastic
modulation of dipole-dipole couplings via their relative motion.
\emph{Intra}molecular NOE enhancements are commonly measured in routine NMR analysis, while the smaller \emph{inter}molecular enhancements, sensitive to average molecular distances, have found applications in studies of intermolecular interactions \cite{Mo1997}.
Similar transfer mechanisms for hyperpolarized systems have been observed in solutions with dissolved ${}^{129}$Xe (dubbed SPINOE for ``spin-polarization-induced nuclear Overhauser effect'' by the authors)\cite{Navon1996,Song:2000} -- as well as with sources polarized via dissolution DNP \cite{Marco-Rius2014} and with parahydrogen induced polarization (PHIP) \cite{Gloeggler:2021}.

In theory, the magnetization transfer from source spins $S$ to target spins $I$ can be described by the semiclassical Solomon equations\cite{Solomon:1955} introducing autorelaxation rates $\rho_{S}\approx1/\rm{T}_{1,S}$ and $\rho_{I}\approx1/\rm{T}_{1,I}$ and cross-relaxation rate $\sigma$ . 

In the situation where the source magnetization is enhanced by a factor $\epsilon_{S}^{\rm{hp}}$ via hyperpolarization, and $\sigma < |\rho_{I}-\rho_{S}|$,\footnote{This is the regime where non-selective relaxometry measurements like the standard T$_1$ inversion recovery with hard pulses yield mono-exponential transient curves with $\rm{T}_{1,I(S)} \approx 1/ \rho_{I(S)}$.} the target enhancement is given by
  \begin{equation}
  \epsilon_{I}^{\rm{hp}}(t) = -\epsilon_{S}^{\rm{hp}} \, \frac{\sigma}{\rho_{I} - \rho_{S}} \left( e^{-\rho_{S}\,t} - e^{-\rho_{I}\,t} \right). 
  \label{eqn:NOE_transient}
\end{equation} 
 
 Note that the magnetization transfer curve follows an initial linear buildup in the opposite direction of the hyperpolarized source magnetization 
\begin{equation} 
  \frac{\text{d}}{\text{d}t}\epsilon_{I}^{\rm{hp}}(t \to 0) = -\epsilon_{S}^{\rm{hp}} \, \sigma. \label{eqn:transient_buildup}
\end{equation} 

Fitting only to eq.\,\ref{eqn:NOE_transient} does not allow the independent determination of the cross-relaxation rate and source polarization, but a standard (thermally polarized) measurement of NOE can be helpful to disentangle both parameters.
In such an experiment, the source spins are saturated by continuous rf irradiation and this leads to a steady-state enhancement of the target spins given as
  \begin{equation} 
  \epsilon_{I}^0 = 1 + \frac{\sigma}{\rho_{I}}.
  \label{eqn:NOE_steady}
\end{equation} 
Separate measurements of T$_1$ then allow $\sigma$ to be determined unambiguously.

\begin{figure*}
  \centering 
  \includegraphics[width=0.8\textwidth]{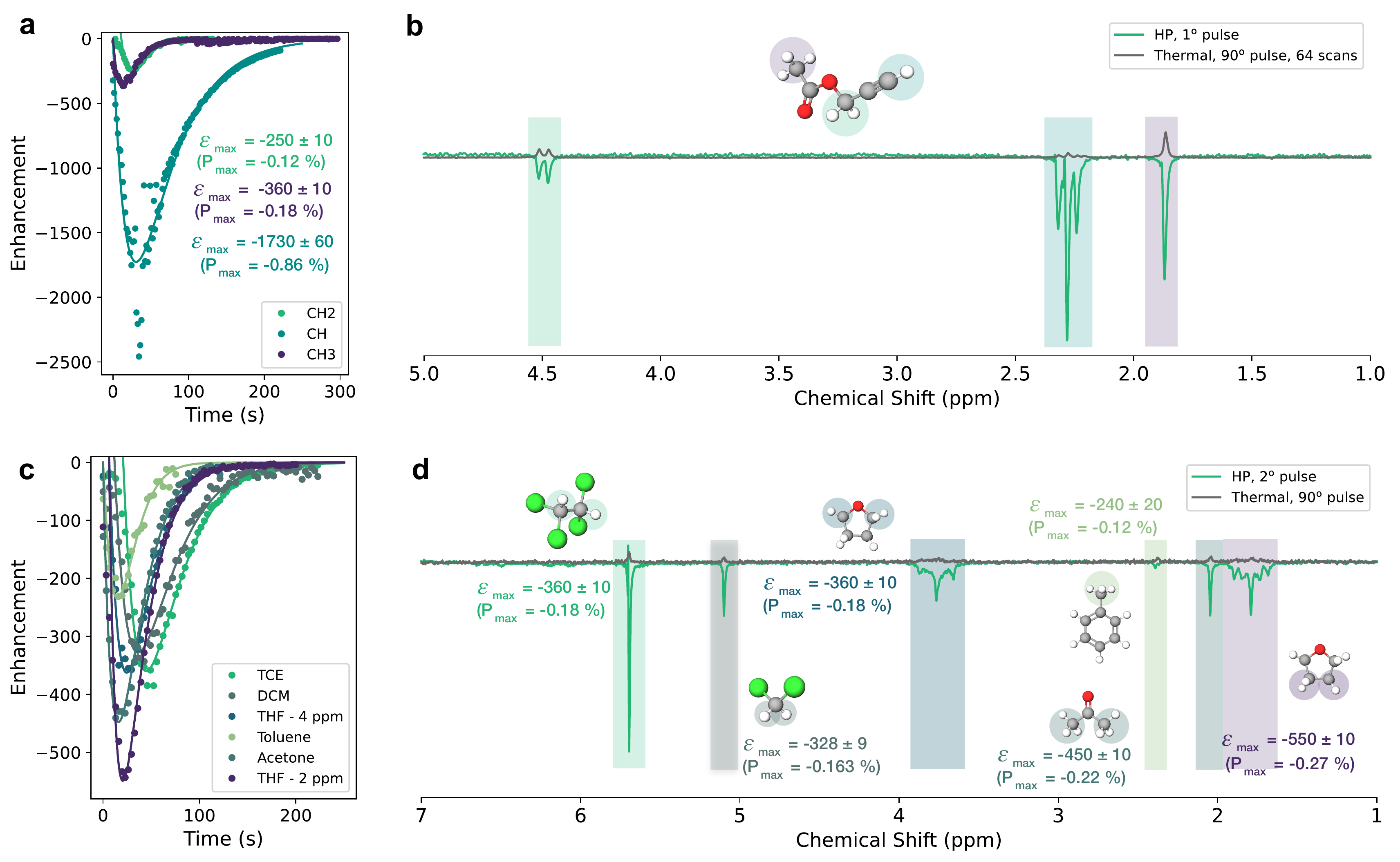}
  \caption{(\textbf{a}) Polarization enhancement over time for the three $^1$H sites of propargyl acetate. 
  Maximum enhancements are reported from fitting the enhancement data to eq.\,\ref{eqn:NOE_transient_modified} (solid lines) such that the values are subject to less error from interference during the measurements.
  (\textbf{b}) $^1$H NMR spectrum acquired with a 1.4$^\circ$ excitation of a hyperpolarized solution containing 100\,mM propargyl acetate 34.4\,s after injection into the spectrometer, compared to the spectrum of the thermally polarized sample acquired with 90$^\circ$ pulses (average of 64 transients). 
  \textbf{c}) Polarization enhancement curves for a solution containing a mixture of small molecules (TCE, DCM, THF, acetone, and toluene).
  \textbf{d}) $^1$H NMR spectrum of the hyperpolarized mixture, acquired with a 2$^\circ$ pulse 22.9\,s after injection into the spectrometer, compared to a spectrum acquired with a 90$^\circ$ pulse after the sample polarization reached thermal equilibrium.}
  \label{fgr:hpspectra}
\end{figure*}

We first consider two model systems: solutions of highly concentrated (1.8\,M, 20\%v/v) naphthalene-$\text{h}_8$ source molecules in CDCl$_3$ containing low concentrations of either acetone or 1,1,2,2-tetrachloroethane (TCE) as a target for spin polarization transfer (the specific compositions of the samples used in Fig.\,\ref{fgr:fig2} are as follows: for Fig.\,\ref{fgr:fig2}a, 100 mM acetone in CDCl$_3$; for Fig\,\ref{fgr:fig2}b, a solvent molecule mixture containing 25 mM acetone in CDCl$_3$, where only the acetone enhancement is considered here; for Fig.\,\ref{fgr:fig2}c, 100\,mM TCE in CDCl$_3$; and for Fig.\,\ref{fgr:fig2}d, 50\,mM TCE in CDCl$_3$). 
While the acetone and TCE concentrations differ between thermal and hyperpolarized experiments, we do not expect this to affect the estimated cross relaxation rate because the naphthalene concentration is much higher than either target concentration, so the source concentration effectively determines the average intermolecular separation in solution.
Both naphthalene and target molecules have T$_1$ relaxation times $>$10\,s, so polarization build-up times are long enough to be easily monitored via small-flip-angle experiments, as shown in Fig.\,\ref{fgr:fig2}. 
Acetone and TCE were chosen as targets here because the T$_1$ relaxation time of acetone, 13.5\,s, is shorter than that of the source, while T$_1$ of TCE, 35.6\,s, is longer.

For thermally polarized samples, the cross-relaxation rate $\sigma$ is extracted according to eq.~\ref{eqn:NOE_steady} using measurements of $\rho_{I}$ and $\epsilon_{I}^0$. 
The insets in figure \ref{fgr:fig2}a,c show the measurement of target thermal $\rm{T}_1=1/\rho$ relaxation times with a standard non-selective inversion recovery sequence, as well as the equilibrium enhancement of magnetization of the target spins after continuously saturating the source spins. 
The normalized cross-relaxation rates evaluated from the thermal saturation measurements are here $\sigma(\rm{acetone}) = (3.7 \pm 1.5) \times 10^{-3}\,\rm{s}^{-1}$ and $\sigma(\rm{TCE}) = (5.6 \pm 0.6) \times 10^{-3} \,\rm{s}^{-1}$,
both of which are significantly higher than values of reported rates between hyperpolarized ${}^{129}$Xe and protons on benzene-d$_5$ and p-nitrotoluene\cite{Song:1997}.
The difference may be explained by the higher nuclear magnetic moment and spin density of the naphthalene $^1$H source as compared to dissolved ${}^{129}$Xe.

Figures \ref{fgr:fig2}b and \ref{fgr:fig2}d show the buildup and decay of $^1$H polarization on acetone and TCE, respectively, at 1.45 T when the naphthalene source has been optically polarized prior to dissolution in the target solution. 

The polarization buildup curves are generally consistent with the characteristic profile expected from eq.~\ref{eqn:NOE_transient}, though there is some notable deviation at short times.
In order to obtain better agreement, the fitting function is modified to include an offset time, $t_0$ and a stretching factor, $\beta$, to the exponential functions:
\begin{equation}
  \epsilon_{I}^{\rm{hp}}(t) = -\epsilon_{S}^{\rm{hp}} \, \frac{\sigma}{\rho_{I} - \rho_{S}} \left( e^{-\rho_{S}(t-t_0)^\beta} - e^{-\rho_{I}\,(t-t_0)^\beta} \right). 
  \label{eqn:NOE_transient_modified}
\end{equation}

The source polarization in solution at the start of the NOE buildup may be estimated via eq.~\ref{eqn:delta_shift}, based on the frequency shifts of the resonances (insets of figures \ref{fgr:fig2}b,d).
This yields $(11.18 \pm 0.06)\,\%$ at $t=0$\,s for the acetone experiment and $(3.461 \pm 0.006)\,\%$ at $t=20$\,s for the TCE experiment.
Fitting to eq.\,\ref{eqn:NOE_transient_modified} (and excluding the first data point) yields 
$\sigma=(3.62 \pm 0.04)\times 10^{-3}\,\rm{s}^{-1}$ for acetone, with zero time offset and stretching factor $\beta=0.980 \pm 0.003$. 
For TCE, the fit (excluding the first ten points) yields $\sigma=(6.24 \pm 0.04)\times 10^{-3}\,\rm{s}^{-1}$ with time offset $(t_0=14.5 \pm0.1)$\,s, and stretching factor $\beta=1.032 \pm 0.001$.
The cross-relaxation rates agree with the thermal measurements, though the physical origins of the time offset and slightly compressed exponential factor for the TCE fit are currently unclear.

\subsubsection{Hyperpolarization of target analytes}

The efficiency of the dissolution NOE polarization transfer approach was benchmarked using a mixture of 20\%v/v naphthalene and 300\,mM propargyl acetate in CDCl$_3$. 
The propargyl acetate spectrum consists of three different proton resonances, all well separated from the naphthalene resonances. 
We show the buildup curves, probed by 1.4$^\circ$ rf pulses, and compare the hyperpolarized spectrum to an averaged, 90$^\circ$ thermal spectrum in Fig. \ref{fgr:hpspectra}a. These curves show the highest enhancement achieved from propargyl acetate, estimated to be  a factor of $-1730 \pm 60$ at 1.45\,T, which corresponds to $0.86\,$\% polarization.
The CH protons exhibit a long $^1$H T$_1$ relaxation time of about 60\,s and a large cross-relaxation rate with naphthalene spins, $(8.4\pm 0.1)\times10^{-3}$\,s$^{-1}$, resulting in a particularly high degree of polarization transfer. 
The CH$_2$ and CH$_3$ resonances both have $^1$H T$_1$s of approximately 10 seconds, as well as lower cross-relaxation rates with naphthalene ($(2.82\pm 0.03)\times10^{-3}$\,s$^{-1}$ and $(3.94\pm 0.06)\times10^{-3}$\,s$^{-1}$, respectively). These target sites reach maximum signal enhancements of $-250 \pm 10$ and $-360 \pm 10$, respectively.

We note that some of the hyperpolarized signals persist for significantly longer than those of thermally polarized samples. 
It is possible that this may be related to positive feedback from the detector back action, along the lines of the so-called ``RASER'' effect \cite{Suefke2017}.
If such effects induce a non-exponential decay, the integrated signals may differ slightly from what would be expected from a small pulse in the absence of feedback.
A more significant complication is that spins may still be precessing by the time that subsequent pulses are applied, which may explain the apparent interference effects in the polarization build-up curve for the -CH resonance in Fig.\,\ref{fgr:hpspectra}a.

We compare the hyperpolarized to thermal signals shown in Fig.\,\ref{fgr:hpspectra}b is chosen at approximately the time where we find the maximum enhancement of the -CH resonance, 34.4\,s after injection to the spectrometer.

In order to demonstrate applicability of the polarization transfer method for more complex NMR spectra, the procedure was applied to a CDCl$_3$ solution containing 100 mM 1,1,2,2-tetrachloroethane (TCE), 100 mM dichloromethane (DCM), 100 mM tetrahydrofuran (THF), 5-10 mM toluene and 25 mM acetone (Fig. \ref{fgr:hpspectra}d,e). 
Signal enhancements greater than 200$\times$ relative to thermal polarization at 1.45\,T were observed for all resonances, between 20 and 40\,s after injection.

\section{Discussion}
In summary, we have shown that the high spin polarization of optically polarized crystals can be transferred to external nuclei by dissolving the polarized material into solutions of interest.
The polarization dynamics are consistent with an intermolecular nuclear Overhauser effect serving as the primary polarization-transfer mechanism.

The large absolute nuclear magnetization (the product of spin density, polarization, and nuclear magnetic moment) achievable with this polarization source--due to the high $^1$H density of naphthalene, the high $^1$H spin polarization from triplet DNP, and the solubility of naphthalene in organic solvents--is the driving force for the level of enhancement observed over this range of molecules. 

Furthermore, this large absolute nuclear magnetization leads to exotic features in the hyperpolarized 1H NMR spectra, including a field drift decaying exponentially over time, as expected due to the geometry of the magnetized sample\cite{Verhulst2002}.
The radiation damping fields are also strong enough to induce nonlinear frequency-dependent phase shifts of the target resonances, as shown in Fig.~\ref{fgr:dataanalysis}.

While the achieved polarization on the target molecules is already substantial, there remains room for improvement. 
First, we estimate a 2- to 5-fold loss in naphthalene polarization through the dissolution process and transfer to the NMR spectrometer, which can likely be at least partially mitigated. 
Additionally, polarizations up to 80\% have been demonstrated\cite{Quan:2019} -- we anticipate that up to 50\% polarization can be achieved with new liquid-nitrogen-cooled systems currently under development.
An order-of-magnitude increase in maximum signal enhancements may therefore be achievable in the future.

Because the polarized crystals in compact permanent magnet assemblies ($\rm{B}_0=0.5\,$T, T = -196$^\circ$ C)\cite{Quan:2019} exhibit $\rm{T}_1\approx 50\,$h, they can be prepared in advance and then transported to the polarization-transfer system.
Thus the optical polarization equipment does not need to be 
at the same facility as the NMR experiments, opening up the prospect of a central facility for production and shipping of highly polarized crystals,\cite{Quan2021Thesis} as has been demonstrated for hyperpolarized Xe and dissolution DNP solutions \cite{Repetto2016,Ji2017}. 

While the experiments in this work focused on small molecules, we anticipate that the method can be applied more generally.
Current efforts are directed to applying the approach to further systems of interest in the areas of
pharmaceutical screening, quality assurance, complex mixture analysis, and structural biochemistry.

\section{Methods}
\subsubsection{Hyperpolarization of source molecules}
The proton spins of naphthalene-h$_8$ are polarized via triplet-DNP\cite{Eichhorn:2013} in a single crystal doped with deuterated pentacene\cite{Eichhorn:2013-2}. 
The photoexcited triplet state of pentacene exhibits a high electron polarization of approx. 90\%, mostly independent of magnetic field and temperature\cite{Strien:1980}. 

Naphthalene is purified by zone-refinement to minimize any contamination, particularly of paramagnetic origin, and maximize the T$_1$ relaxation time and hence triplet-DNP yield. 
The quality of the material is periodically probed during the purification process by means of delayed fluorescence\cite{Wakayama:1982}. 
Pentacene-doped naphthalene crystals are grown with the self-seeding Bridgman method\cite{Selvakumar:2005} and pentacene doping concentrations are preferentially chosen close to saturation found in the order of 10 - 100 ppm to maximize triplet-DNP buildup rates.
We achieve triplet-state lifetimes exceeding 300$\,$ms in pentacene-doped naphthalene at room temperature, indicating relatively high levels of purity. 

For the experiments performed here, samples are cut to approx. 40 mg and the crystalline axes are identified by optical birefringence to facilitate sample alignment.
The sample is mounted in a home-built triplet-DNP system, which consists of a cryostat, equipment for optical excitation of triplet states and crystal axis alignment, a microwave cavity, an electromagnet, and NMR electronics.

Cooling uses a jet of evaporated liquid nitrogen to reach temperatures around -150°C. 
Photoexcited triplet states are initialized with a pulsed laser (527 nm, ca. 100 ns pulse length, typically 500 Hz to 1 kHz operation at ca. 500 mW average power)
 and the electron spin polarization of the triplet state is optimized by aligning the sample to its canonical orientation (B$_0$ || pentacene X-axis).
 
To transfer polarization from the optically polarized electron spins to 1H nuclei in the pentacene-naphthalene crystal, the sample is shuttled into a microwave cavity in the field of an electromagnet maintained at 0.22 T, where an optimal-control enhanced transfer scheme sweeping microwave frequency and phase is applied with an arbitrary waveform generator. The initial guess of the optimization algorithm was a `standard' integrated solid effect sweeping the microwave frequency applied to the cavity \cite{Henstra1988b}.

For NMR readout of the nuclear spin polarization, the sample is shuttled to a 0.8T NMR probe just above the cavity. The polarization is calibrated by comparison of the hyperpolarized signal to that of a thermally polarized pentacene-doped naphthalene crystal calibration standard.
Control, readout and analysis of the system is performed with the software suite qudi\cite{binder:2017}. 

Under optimized conditions, the homebuilt triplet DNP system achieves polarization buildup rates ranging typically between 0.1 and 1\%/min (dependent on doping concentration and sample thickness) levelling between 20 and 25\,\% on average for 40 mg samples. Given the high absorption cross section, we typically assumed the crystal to be inhomogeneously polarized along the laser direction.

\subsubsection{Thermal NMR measurements}

In the experiments we use two benchtop spectrometers, one which is included in the polarization transfer setup (1.45 T) to monitor polarization transfer and thermal calibration, and one which we use to perform the saturation transfer measurements to gauge cross-relaxation rates (1.88 T).  

Intermolecular NOE enhancements were obtained via a difference measurement where the enhanced signal is compared to a reference signal acquired with the same pulse sequence but with the saturation frequency set at an equal offset opposite to the target resonance\cite{kaiser1963use}.

\subsubsection{Polarization transfer to target molecules}
In order to bring the polarization source in contact with the target molecules of choice, a polarized crystal is transported via a handheld magnet assembly (B$_0 \approx 150$\,mT) from the optical polarizer to a ceramic crushing and mixing vessel. 
The crystal rests on a titanium frit with pore size on the order of $5\,\mu$m. 
In order to maximize the polarization transfer efficiency we aim to reduce the average distance of source and target molecules as much as possible and therefore dissolve naphthalene to 1.6-1.8 M which is not far from its saturation point in chloroform at room temperature. 
To achieve concentrations near 20\%(v/v) naphthalene, approximately 160\,$\mu$L of target solution are injected into the vessel.

The dissolution of the crystal is achieved by the lowering and rotation of a motor-controlled ceramic shaft into the ceramic vessel. 
Crushing and mixing occurs for 6 seconds until the majority of the polarized naphthalene crystal is dissolved. The extent of naphthalene dissolution with this procedure was verified prior to hyperpolarized experiments using unpolarized crystals and calibrating on the thermal naphthalene signal in CDCl$_3$.

The material is then pushed through the titanium frit and injected over the course of 8 seconds to a 3 mm NMR tube positioned into a benchtop NMR spectrometer for monitoring the hyperpolarized signals over time. 
The hyperpolarized signals are monitored with small-angle hard-pulse excitations typically of 1-3$^\circ$ to minimize effects of radiation damping, but still provide adequate SNR of the lower concentration target molecule signals. 
Pulses are spaced either by 1.7 or 3.3 second acquisition periods.

In the experiments presented in this manuscript, the following target solutions were used: (1) 100 mM  or 300 mM propargyl acetate (Sigma Aldrich) in 99.8\% D CDCl$_3$, (2) 50 mM TCE in 99.8\% D CDCl$_3$, (3) (2) 100 mM DCM in 99.8\% D CDCl$_3$, or a mixture of small molecules: (4) 25 mM acetone, 100 mM dichloromethane, 160 mM Chloroform, 100 mM THF, 100 mM TCE, and 10 mM toluene in 99.8\% D CDCl$_3$. 
Each target solution was bubbled with N$_2$ gas for 120 s before each experiment to expel any paramagnetic O$_2$ in order to minimize sources of nuclear spin autorelaxation.
An inert atmosphere of N$_2$ gas was also maintained during the process of crushing/dissolving the crystal, and pressurized N$_2$ gas was used to inject the hyperpolarized mixture to the spectrometer.



\bibliography{references.bib}

\end{document}